\documentclass[conference]{IEEEtran}
\IEEEoverridecommandlockouts
\usepackage{cite}
\usepackage{amsmath,amssymb,amsfonts}
\usepackage{algorithmic}
\usepackage{graphicx}
\usepackage{booktabs}
\usepackage[hidelinks]{hyperref}
\usepackage{textcomp}
\usepackage{url}
\usepackage{xcolor}
\def\BibTeX{{\rm B\kern-.05em{\sc i\kern-.025em b}\kern-.08em
    T\kern-.1667em\lower.7ex\hbox{E}\kern-.125emX}}
\begin{document}

\title{Benchmarking miRNA Target Prediction Methods }
\author{
\IEEEauthorblockN{Congliang Jin\textsuperscript{1},
Weibo Niu\textsuperscript{1},
Zhexing Wen\textsuperscript{1},
Zhaohui S Qin\textsuperscript{1},
Ximing Ran\textsuperscript{1,*}}
\IEEEauthorblockA{\textsuperscript{1}Emory University, Atlanta, GA, USA\\
\textsuperscript{*}Corresponding author: ximing.ran@emory.edu}
}

\maketitle

\begin{abstract}
\textbf{Background:}
MicroRNAs (miRNAs) are small non-coding RNAs that post-transcriptionally regulate gene expression by binding to target messenger RNAs, making accurate identification of their targets essential to understanding miRNA-mediated regulation. Although numerous computational tools predict miRNA--target interactions, their predictions show limited concordance, and existing benchmarks often rely on curated databases that may overlap with the data used to develop these tools, potentially biasing evaluation.
\par\noindent\textbf{Methods:}
We benchmarked seven widely used tools (TargetScan, miRDB, miRmap, DIANA-microT, RNA22, miRWalk, and mirDIP) on approximately 59.7 million candidate miRNA--mRNA pairs. Each pair was assigned an evidence-graded reference score (0--3) integrating experimentally validated interactions from two external databases (miRTarBase and TarBase) with expression-derived evidence from paired miRNA-seq and mRNA-seq profiles of an internal induced pluripotent stem cell (iPSC) cohort comprising 22q11.2 deletion syndrome and control lines. Performance was assessed by Type I error rate, statistical power, and false discovery proportion (FDP). Robustness was examined by increasing the stringency of the reference-score cutoff (truth signal) and by progressively admitting lower-confidence predictions (model signal). Optimal score cutoffs were selected by maximizing Youden's $J$ statistic.
\par\noindent\textbf{Results:}
Tool rankings were largely consistent across evaluation settings. TargetScan and miRDB maintained Type I error rates below 0.05 in all settings, with TargetScan yielding the lowest FDP, but both had limited power. mirDIP achieved the highest power at the cost of the highest Type I error rate. At optimal cutoffs, mirDIP provided the best balance between sensitivity and specificity (Youden's $J = 0.566$), followed by DIANA-microT ($J = 0.481$), whereas miRWalk and RNA22 had FDPs close to the random-prediction baseline. These findings indicate that no single tool is optimal across all metrics, and that tool and cutoff selection should be guided by whether a study prioritizes sensitivity or specificity.
\end{abstract}


\begin{figure*}[!t]
    \centering
    \includegraphics[
        width=\linewidth,]{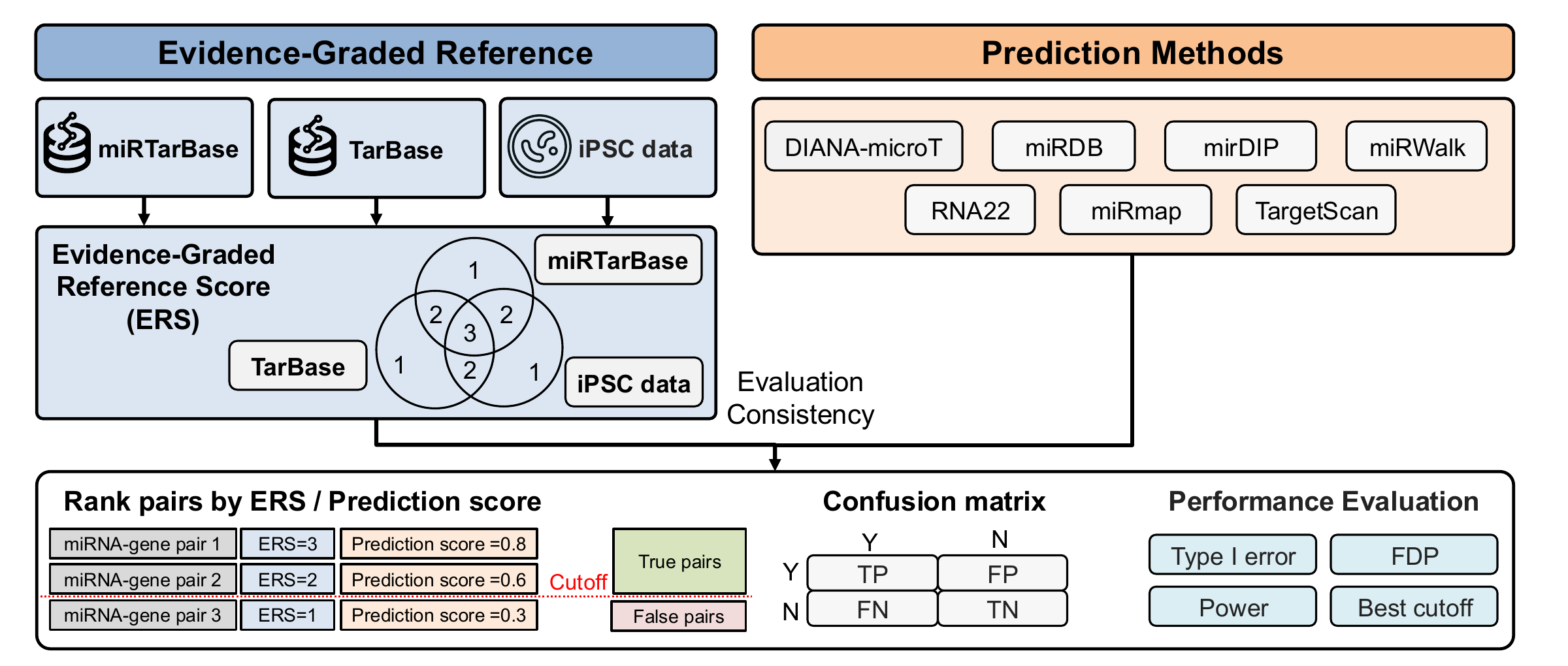}
    \caption{\textbf{Overview of the benchmarking design.} Paired miRNA-seq and mRNA-seq data from the internal iPSC dataset are integrated with miRTarBase and TarBase to construct a reference scoring system (0--3). Seven prediction tools are evaluated using Type I error rate, statistical power, and FDP across varying reference-score cutoffs (truth signal) and prediction confidence thresholds (model signal).}
    \label{fig1}
\end{figure*}


\section{Introduction}
MicroRNAs (miRNAs) are small non-coding RNAs of $\sim$22 nucleotides that regulate gene expression by binding to target mRNAs, leading to translational repression or mRNA degradation. Accurate identification of miRNA targets is essential for understanding gene regulation and disease mechanisms. Because experimental validation is labor-intensive, various computational prediction tools have been developed using features such as sequence complementarity, binding energy, and evolutionary conservation, as well as machine learning approaches.

Despite the growing number of miRNA target prediction tools, selecting an appropriate method remains challenging due to limited agreement among their predictions. Existing benchmarks often evaluate only a subset of tools, rely on heterogeneous references, or use data overlapping with tool training sets, potentially introducing evaluation bias. Moreover, experimentally validated databases remain incomplete, and tool performance in induced pluripotent stem cells (iPSCs) is less explored.

In this study, we benchmark seven widely used miRNA target prediction tools (TargetScan, miRDB, miRmap, DIANA-microT, RNA22, miRWalk, and mirDIP) using paired miRNA-seq and mRNA-seq data from our internal iPSC dataset, derived from donors with 22q11.2 deletion syndrome (22qDS) and matched controls, together with two external experimentally validated databases, miRTarBase and TarBase.

Rather than evaluating prediction tools against a single fixed reference, we introduce a statistical benchmarking framework that combines experimentally validated miRNA--gene interactions with expression-derived evidence. Each candidate pair receives a reference score (0--3) reflecting how many of the three evidence sources support it. We evaluate seven prediction tools using Type I error rate, statistical power, and FDP, and assess robustness by varying the reference-score cutoff (truth signal) and prediction confidence threshold (model signal).

The main contributions of this paper are:

\begin{itemize}
    \item \textbf{Evidence-graded reference from three datasets.} Rather than treating the reference as a single binary standard, we evaluate each candidate miRNA--target pair by how many of the three evidence sources support it, including paired miRNA and mRNA expression profiles from an internal iPSC cohort with 22q11.2 deletion syndrome and matched controls. This reduces reliance on curated databases alone and places the comparison in a biologically relevant setting.

    \item \textbf{Cutoff evaluation across evidence levels.} For each tool, we evaluate prediction score cutoffs at each evidence level (pairs supported by at least one, at least two, or all three datasets) and report Type I error rate, power, and false discovery proportion (FDP) jointly. We then select each tool's optimal cutoff using Youden's $J$ statistic, which balances Type I error and power.

    \item \textbf{Best-practice recommendations.} Comparing tools at their optimal prediction score cutoffs shows that no single tool is best in every setting: the right choice depends on whether a study prioritizes high power or a low FDP. We summarize these trade-offs as practical guidance for choosing both the tool and its cutoff.
\end{itemize}

\section{Related Work}
Seven tools were selected, each cited more than 200 times since 2020 and providing genome-wide predictions for direct download (Table~\ref{tab:tools}). For each tool, only predictions for human miRNAs present in miRBase~v22 were kept; where multiple entries existed for the same miRNA--mRNA pair, only the best-scoring entry was retained. Targets were ranked by each tool's native interaction score to enable score-quantile analyses. Together, the seven tools cover four major methodological categories and remain among the most
widely used publicly available predictors. In addition, miRTarBase and
TarBase were included as experimentally validated databases to provide
reference evidence for evaluating the prediction tools.

Thermodynamic-based tools (PITA, RNAhybrid) and deep-learning approaches were excluded: the former lack publicly available genome-wide predictions for direct download, and the latter offer limited interpretability in the high-dimensional, small-sample setting of this cohort. Both nonetheless represent promising directions for future benchmarking.

\begin{table}[h!]
\caption{\textup{miRNA} target prediction tools and two validated databases.}
\label{tab:tools}
\centering
\footnotesize
\renewcommand{\arraystretch}{1.18}
\setlength{\tabcolsep}{3pt}
\newcommand{\tworow}{\vphantom{\parbox[t]{0pt}{\strut\\\strut}}}

\begin{tabular}{@{}p{2.5cm}p{1.45cm}p{4.45cm}@{}}
\toprule
\textbf{Tool/Database} & \textbf{Category} & \textbf{Key features} \\
\midrule

\tworow TargetScan \cite{targetscan}
& Seed/rule-based
& 7mer/8mer seed matching; conservation;
$P_{\mathrm{CT}}$; context++ scoring \\

\tworow miRDB \cite{mirdb}
& Machine learning
& SVM-based prediction; sequence
and expression features \\

\tworow miRmap \cite{mirmap}
& Feature-scoring
& Binding thermodynamics ($\Delta G$);
accessibility; AU content; phyloP \\

\tworow DIANA-microT \cite{dianamicrot}
& Feature-scoring
& MRE features; binding energy;
evolutionary conservation \\

\tworow RNA22 \cite{rna22}
& Seed/rule-based
& Pattern-based sequence matching;
flexible seed pairing; target islands \\

\tworow miRWalk \cite{mirwalk}
& Integration
& 5$'$UTR, CDS, and 3$'$UTR regions \\

\tworow mirDIP \cite{mirdip}
& Integration
& Confidence-based ranking \\

\tworow miRTarBase \cite{mirtarbase}
& Validated database
& Curated experimentally validated
miRNA--target interactions \\

\tworow TarBase \cite{tarbase}
& Validated database
& Experimentally supported
miRNA--target interactions \\

\bottomrule
\end{tabular}
\end{table}

\begin{figure*}[!t]
    \centering
    \includegraphics[width=\linewidth]{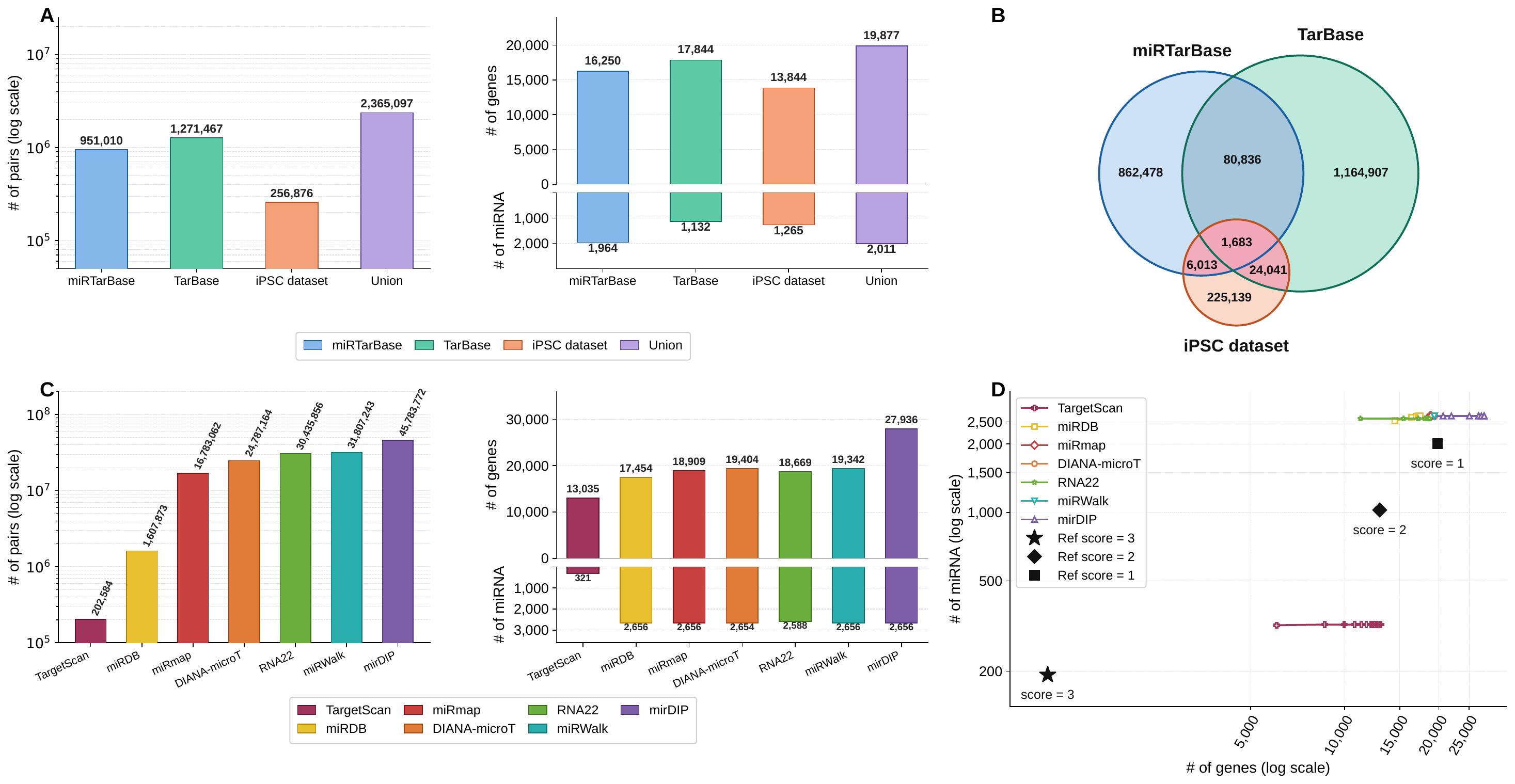}
    \caption{\textbf{Reference consistency and tool coverage.} (A)~Interaction pairs, genes, and miRNAs across reference datasets. (B)~Overlap among the three reference datasets. (C)~Prediction coverage of seven tools. (D)~Centrality of each tool, showing gene and miRNA coverage across prediction thresholds (top 10\%--100\%) and reference-score cutoffs.}
    \label{fig2}
\end{figure*}
\section{Methods}

\subsection{Internal iPSC dataset and preprocessing}
The 22q11.2 deletion syndrome (22qDS) is the most prevalent human microdeletion disorder, affecting approximately 1 in 3{,}000 to 1 in 6{,}000 live
births and an estimated 1 in 1{,}000 fetuses. The deleted region includes \textit{DGCR8}, which mediates miRNA biogenesis~\cite{mcdonaldmcginn2015}. We used an iPSC cohort comprising 5~control and 5~22qDS donor lines, each sequenced in 2 technical replicates, yielding 20~matched mRNA-seq and small RNA-seq libraries. mRNA-seq reads were aligned to GRCh38 (Ensembl release 106) with HISAT2 v2.2.1~\cite{hisat2} and quantified with featureCounts v2.0.2~\cite{featurecounts} in paired-end, stranded mode; small RNA-seq reads were processed with nf-core/smrnaseq v2.3.0, including adapter trimming and size filtering to 18--26~nt, and quantified against miRBase v22 using miRDeep2~\cite{mirdeep2}. Counts were normalized with the DESeq2 median-of-ratios method and variance-stabilized~\cite{deseq2}, technical replicates were averaged on the VST scale, and donor and batch effects were removed with ComBat-seq~\cite{combatseq}, conditioned on 22qDS status. miRNAs with mean expression below the LOESS-estimated inflection point of the standard-deviation curve were excluded, leaving 29,339 expressed mRNA
genes and 2,036 miRNAs and yielding 59,734,204 candidate miRNA--mRNA pairs.

\subsection{Reference construction}
As shown in Fig.~1, the reference integrates two external experimentally validated databases (miRTarBase and TarBase) with internal expression-derived evidence from paired miRNA-seq and mRNA-seq profiles (the \textit{iPSC} dataset, Fig.~2A). Expression support required a negative Pearson correlation ($r < -0.2$) and a Benjamini--Hochberg adjusted FDR $< 0.05$. Each candidate pair received a reference score from 0 to 3, one point per supporting source, and was labeled reference-positive if its score met a cutoff $c \in \{1, 2, 3\}$ and reference-negative otherwise. Raising $c$ restricts positives to pairs with stronger evidence while keeping all candidate pairs in the evaluation, which lets us test whether performance holds under stricter reference definitions. 

\subsection{Type I error evaluation}
\textbf{Type I error evaluation assesses how frequently each tool predicts interactions among reference-negative pairs.} The Type I error rate is the proportion of reference-negative pairs predicted as positive $(FP/(FP+TN))$, benchmarked against the conventional $\alpha = 0.05$. In the truth-signal analysis (Fig.~3A), we used each tool's full prediction set and varied $c$ from 1 to 3. In the model-signal analysis (Fig.~3B), we fixed $c=1$ and retained the top 10\%, 20\%, $\ldots$, 100\% of predictions ranked by prediction score. A lower rate indicates better false-positive control.

\subsection{Statistical power evaluation}
\textbf{Power evaluation assesses how completely each tool recovers reference-positive interactions.} Power is the proportion of reference-positive pairs correctly predicted $(TP/(TP+FN))$, i.e., one minus the Type II error rate ($1-\beta$), benchmarked against a target power of 0.80. We used the same truth-signal and model-signal analyses as above (Fig.~4A and 4B). Higher power indicates a lower Type II error rate.

\subsection{False discovery evaluation}
\textbf{False discovery evaluation assesses the agreement between each tool's predictions and the reference.} The FDP is the proportion of predicted-positive pairs that are reference-negative $(FP/(TP+FP))$; unlike power, it is computed over predicted positives. In the truth-signal analysis (Fig.~5A), we retained each tool's top 20\% of ranked predictions to focus on high-confidence candidates and varied the reference-score cutoff $c$ from 1 to 3. In the model-signal analysis (Fig.~5B), we fixed $c=1$ and progressively retained the top 10\%--100\% of each tool's ranked predictions. The dashed line in Fig.~5B indicates the expected FDP under random prediction, equal to the proportion of reference-negative pairs among all candidates at $c=1$ ($57{,}369{,}107/59{,}734{,}204 \approx 0.96$). A lower FDP indicates more reliable predictions.

\newpage
\section{Results}

\subsection{Coverage and overlap of reference sources}

The three reference sources were largely complementary, with little overlap. TarBase contributed the most interaction pairs (1,271,467), followed by miRTarBase (951,010) and the expression-derived dataset (256,876) (Fig.~2A). Their union, which defines the reference-positive set at $c=1$, comprised 2,365,097 unique pairs spanning 19,877 genes and 2,011 miRNAs, yet only 1,683 pairs were supported by all three sources (Fig.~2B).

\subsection{Coverage of prediction tools}

Prediction coverage varied widely across the seven tools. mirDIP covered the most interaction pairs, whereas TargetScan covered relatively few (Fig.~2C), and gene and miRNA coverage also differed across prediction-score thresholds (Fig.~2D). Relaxing the reference cutoff from $c=3$ to $c=1$ expanded reference coverage, as expected.

\subsection{Type I error comparison}

TargetScan and miRDB showed the strongest false-positive control, whereas mirDIP showed the weakest. In the truth-signal analysis, TargetScan and miRDB stayed below the 0.05 benchmark at all cutoffs, mirDIP had the highest Type I error rate, and miRWalk and RNA22 were also relatively high, with the tool ranking stable across cutoffs (Fig.~3A). In the model-signal analysis, Type I error rates rose as more predictions were included; TargetScan and miRDB remained low even with all predictions retained, whereas mirDIP increased sharply (Fig.~3B).

\begin{figure}[!h]
    \centering
    \includegraphics[width=\linewidth]{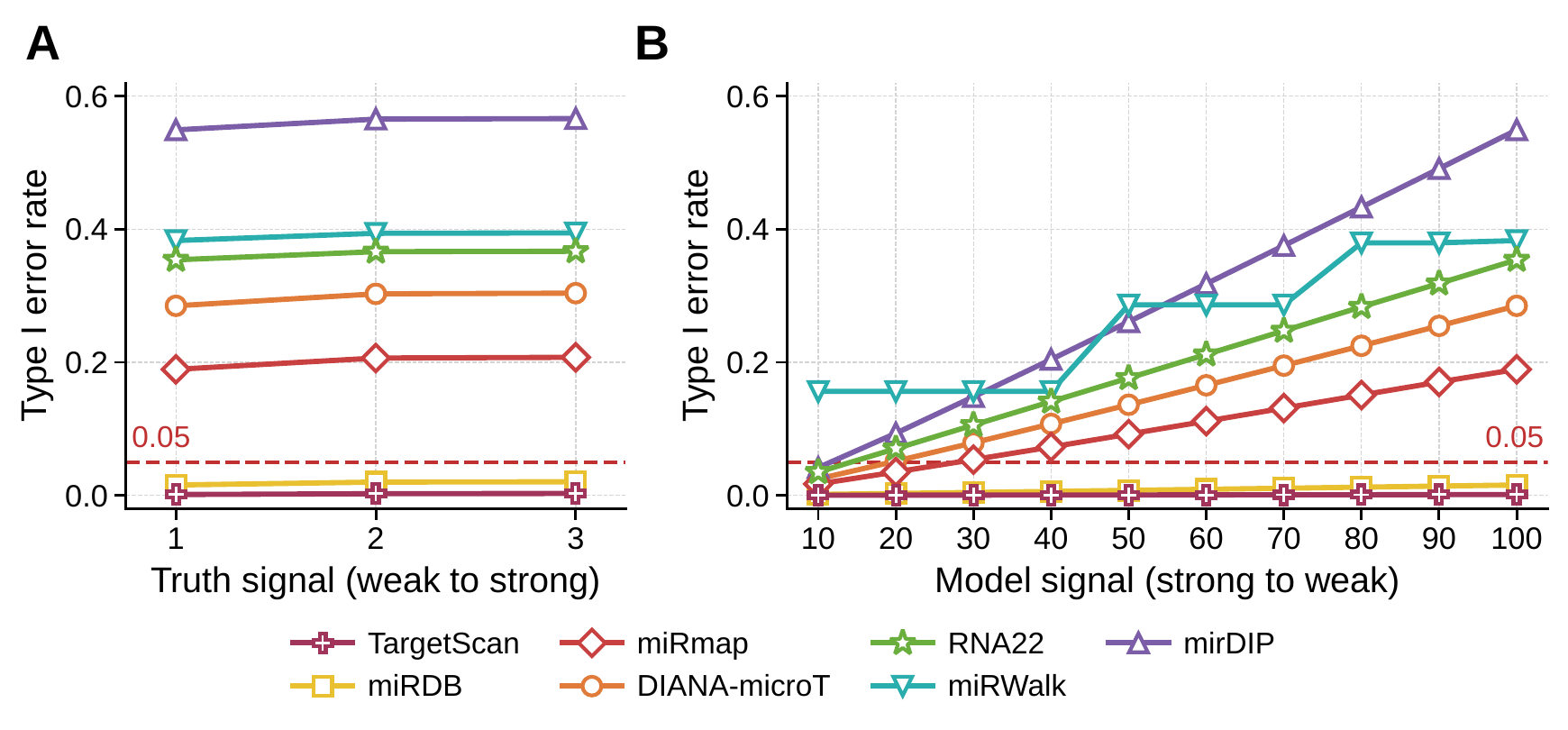}
    \caption{\textbf{Type I error rate of the seven tools (lower is better).}
    (A) Type I error rates across reference-score cutoffs ($c=1,2,3$). (B) Type I error rates across prediction thresholds (top 10\%--100\%) at $c=1$.
    The dashed line indicates the 0.05 benchmark.}
    \label{fig3}
\end{figure}

\subsection{Statistical power comparison}

mirDIP achieved the highest power at the cost of the highest Type I error rate, whereas TargetScan and miRDB showed the opposite pattern. In the truth-signal analysis, mirDIP's power stayed close to 1.0 at all cutoffs, and DIANA-microT and miRmap exceeded the 0.80 benchmark at stricter cutoffs (Fig.~4A). In the model-signal analysis, mirDIP exceeded 0.80 with the top 50\% of predictions retained, while no other tool reached the benchmark even with all predictions included (Fig.~4B).
\begin{figure}[!h]
    \centering
    \includegraphics[width=\linewidth]{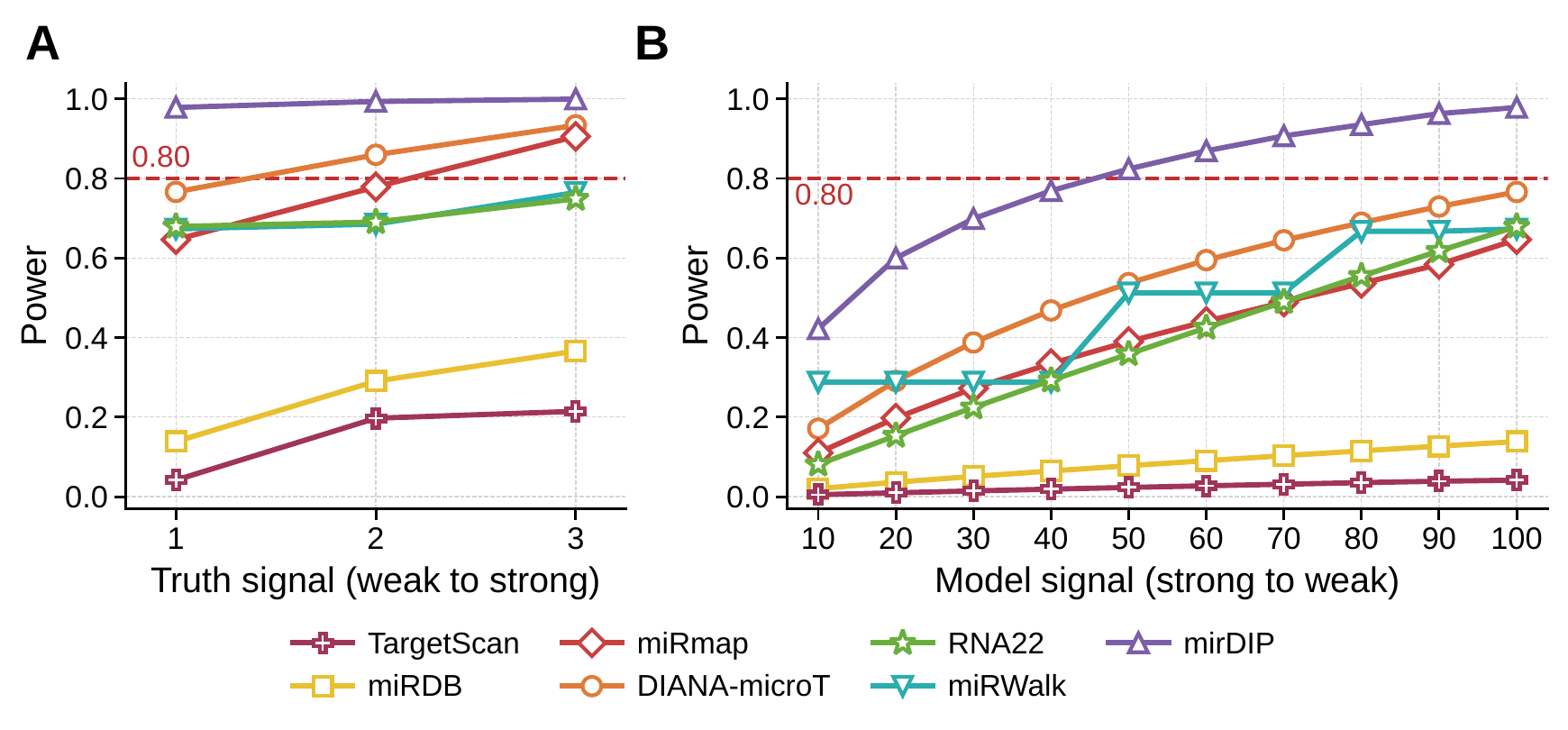}
    \caption{\textbf{Statistical power comparison across seven prediction tools (higher is better).}
    (A) Statistical power across reference-score cutoffs ($c=1,2,3$).
    (B) Statistical power across prediction thresholds (top 10\%--100\%) at $c=1$. The dashed line indicates the 0.80 benchmark.}
    \label{fig4}
\end{figure}

\subsection{False discovery comparison}

TargetScan produced the most reliable predictions, whereas miRWalk and RNA22 performed close to random. In the truth-signal analysis, FDP rose for all tools as the cutoff became stricter, approaching 1.0 for most tools at $c=3$; TargetScan rose the most, from about 0.30 to nearly 1.0 (Fig.~5A). In the model-signal analysis, FDP increased gradually as more predictions were retained, with TargetScan lowest at every threshold, followed by miRDB, while miRWalk and RNA22 stayed near the random baseline of 0.96 (Fig.~5B).

\begin{figure}[!h]
    \centering
    \includegraphics[width=\linewidth]{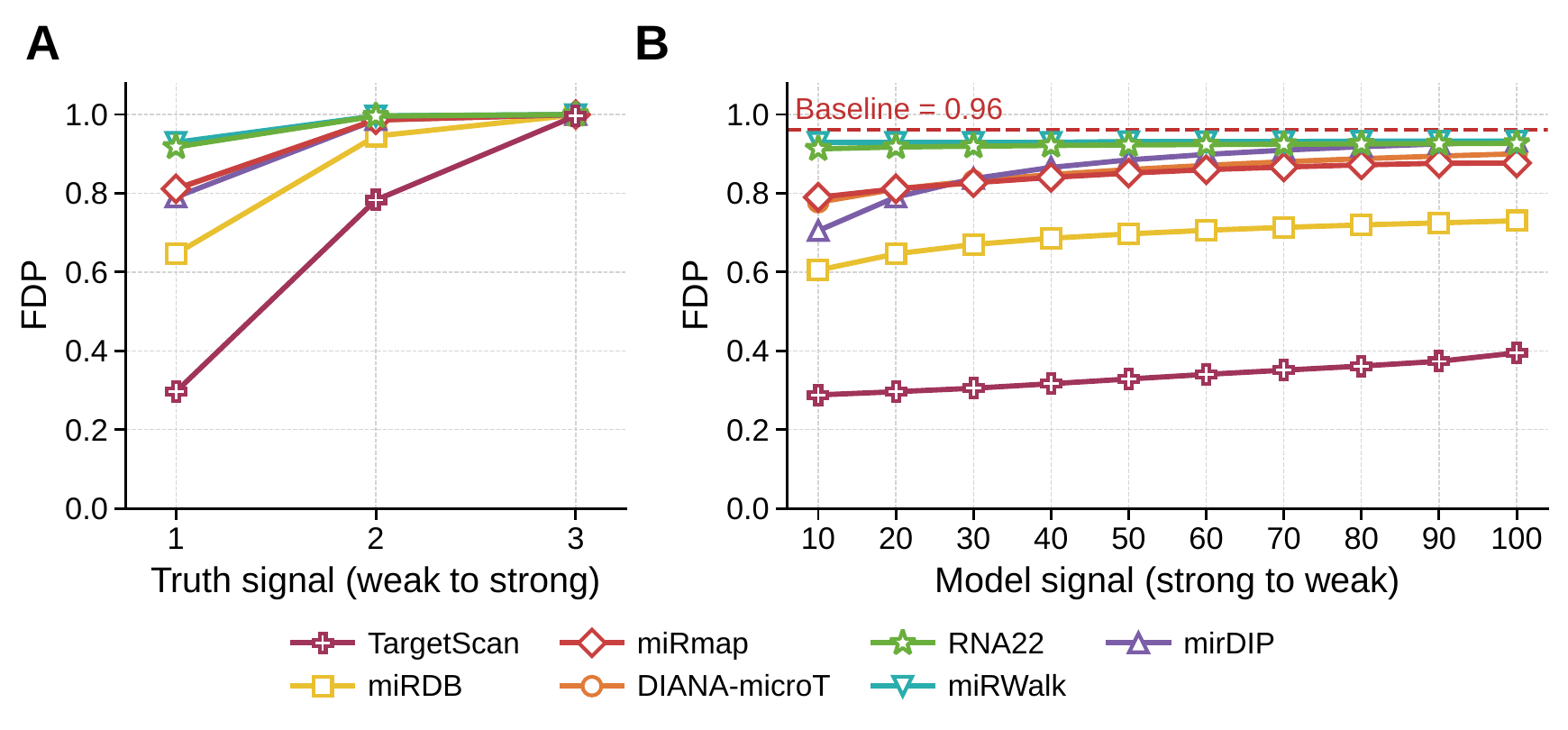}
    \caption{
        \textbf{False discovery proportion (FDP) comparison across seven prediction tools (lower is better).}
        (A) FDP across reference-score cutoffs ($c=1,2,3$) as the truth signal strengthens.
        (B) FDP across prediction thresholds (top 10\%--100\%) as the model signal weakens, with the reference-score cutoff fixed at $c=1$.
The dashed line indicates the random-prediction baseline (FDP = 0.96).
    }
    \label{fig5}
\end{figure}

\section{Discussion and Conclusions}

This study systematically compared seven miRNA target prediction tools
using paired miRNA and mRNA expression data from induced pluripotent stem cells (iPSCs), covering approximately 59.7 million candidate miRNA--mRNA interactions. Tool rankings were largely consistent across evaluation settings, but no single tool performed best on all metrics. mirDIP achieved the highest Youden's $J$ (0.566) at its optimal cutoff, followed by DIANA-microT (0.481; Table~\ref{tab:error_rates}), indicating the strongest discrimination between reference-positive and reference-negative interactions.

These findings highlight that tool selection should follow the research objective. Based on our results, we suggest the following \textbf{best practices}:

\begin{itemize}
    \item \textbf{Comprehensive screening.} When the goal is to capture as many potential interactions as possible, mirDIP is preferred. At its best cutoff ($\geq 0.061$), it recovered 80\% of reference-positive interactions, but with a Type I error rate of 0.235, so its output should be treated as a candidate list for further filtering.
    \item \textbf{High-confidence candidates for experimental validation.} TargetScan and miRDB are preferred, as they were the only tools that kept Type I error below 0.05 in every setting and had the lowest FDP. However, they recovered only 4\%--14\% of reference-positive interactions at their best cutoffs, so the absence of a prediction should not be taken as evidence of no interaction.
    \item \textbf{Balanced use.} mirDIP and DIANA-microT offer the best overall trade-off, while miRmap provides a lower Type I error rate (0.189) at a moderate cost in power.
    \item \textbf{Score cutoff selection.} Restricting to the top 10\% of predictions kept Type I error below 0.05 for all tools except miRWalk, at the cost of power, so cutoffs should be chosen and reported according to the tolerance for false positives.
    \item \textbf{Corroborating evidence.} Even allowing for the
    incompleteness of the reference, five of the seven tools had FDP above 0.85 at their best cutoffs, and miRWalk and RNA22 performed close to the random baseline. Predictions should therefore be combined with independent evidence, such as experimentally validated databases or context-specific expression data, rather than used alone.
\end{itemize}

\begin{table}[!htbp]
\centering
\caption{Error rates of each method at its best score cutoff.}
\label{tab:error_rates}
\footnotesize
\renewcommand{\arraystretch}{1.15}

\resizebox{\columnwidth}{!}{%
\begin{tabular}{@{}lccccc@{}}
\toprule
\textbf{Method} &
\textbf{Best cutoff} &
\textbf{FDP} &
\textbf{Type I error} &
\textbf{Power} &
\textbf{Youden's J} \\
\midrule
TargetScan   & $\leq 0.000$  & 0.395 & 0.001 & 0.042 & 0.040 \\
miRDB        & $\geq 50.000$ & 0.730 & 0.015 & 0.138 & 0.123 \\
miRmap       & $\leq 8.155$  & 0.877 & 0.189 & 0.646 & 0.457 \\
DIANA-microT & $\geq 0.352$  & 0.900 & 0.283 & 0.765 & 0.481 \\
RNA22        & $\leq 0.390$  & 0.927 & 0.354 & 0.679 & 0.325 \\
miRWalk      & $\geq 0.801$  & 0.932 & 0.383 & 0.673 & 0.290 \\
mirDIP       & $\geq 0.061$  & 0.877 & 0.235 & 0.800 & 0.566 \\
\bottomrule
\end{tabular}%
}
\end{table}

Integrating miRTarBase, TarBase, and expression-derived evidence reduces reliance on any single database, and tool rankings remained stable across settings. No tool achieved both high power and strict false-positive control, so performance should be judged with multiple metrics. Because the three reference datasets may not fully represent the true regulatory landscape and the expression data come from a single 22qDS iPSC cohort, results should be interpreted with caution. Our future work will integrate the miRNA--gene relationships captured by widely used tools and databases into a unified framework and extend this benchmark to other cellular and disease contexts.

\section*{ Data Availability:}
The source code is publicly available at \url{https://github.com/congliangjin141-coder/miRNA_target_prediction}.

\section*{Acknowledgment}
This work was supported by the NARSAD Young Investigator Grant from the Brain \& Behavior Research Foundation (BBRF) to W.N. (James John Karatheodoris Memorial Investigator, Grant No.~33517).

\bibliographystyle{IEEEtran}
\bibliography{literature}
\end{document}